\documentstyle[12pt]{article}
\newcommand{\prd}[3]{{\it Phys. Rev.} {\bf D#1}, #2 (19#3)}
\newcommand{\pl}[3]{{\it Phys. Lett.} {\bf #1B}, #2  (19#3)}
\newcommand{\np}[3]{{\it Nucl. Phys.} {\bf B#1}, #2  (19#3)}
\newcommand{\prl}[3]{{\it Phys. Rev. Lett.} {\bf #1}, #2  (19#3)}
\newcommand{\ra}{\rightarrow}
\newcommand{\matel}[3]{\left<#1\right|#2\left|#3\right>}
\newcommand{\cp}{{\cal CP}}

\def\uu{\overline{u}}
\def\dd{\overline{d}}

\def\ll{\lambda^a}
\def\gm{\gamma_\mu}

\relax
\begin{document}
\def\ba{\begin{array}}
\def\ea{\end{array}}
\def\thefootnote{\fnsymbol{footnote}}
\vspace{0.5in}
\hskip 4in ISU-HET-95-8
\vspace{0.2in}
\begin{center}
{\large \bf Theoretical Aspects of $CP$ Violation in Hyperon Decays}\\
\vspace{0.2in}
{\bf  G.~Valencia}\\
{\it           Department of Physics and Astronomy,
               Iowa State University,
               Ames, IA 50011}\\
\end{center}

\vspace{0.2in}
\begin{abstract}

\noindent
I review the estimate of the $CP$ violating asymmetry $A(\Lambda^0_-)$
within the standard model. I then review the estimate of the upper bound
on this asymmetry that arises from measurements of $CP$ violation in
kaon decays.

\end{abstract}
\vspace{0.2in}

\noindent
One of the systems where it is possible to search for $\cp$ violation is in
the non-leptonic decay of hyperons.
Of particular interest for  the upcoming experiment E871 is the asymmetry
$A(\Lambda^0_-)$ \cite{donpa}. E871
expects to reach a sensitivity of $10^{-4}$ for the sum of asymmetries
$A(\Lambda^0_-)+ A(\Xi^-_-)$ \cite{proposal}.
Unfortunately, the calculation of these asymmetries is
plagued by theoretical uncertainties in the estimate of the hadronic matrix
elements involved. Nevertheless, a conservative study of these asymmetries
within the minimal standard model has shown that $A(\Lambda^0_-)$ is likely to
occur at the level of a few times $10^{-5}$ \cite{steger}. Similarly, a recent
estimate has indicated that current constraints from measurements of $\cp$
violation in kaon decays do not preclude $A(\Lambda^0_-)$ from reaching values
of order a few times $10^{-4}$ beyond the standard model \cite{latest}.
In view of this, the potential
results of E871 are very exciting.

The general framework to discuss $A(\Lambda^0_-)$ can be found for
example in Ref.~\cite{donpa}. With the experimentally known values for
the strong rescattering phases and the moduli of the weak decay
amplitudes, we can write:
\begin{equation}
A(\Lambda^0_-)\approx 0.13 \sin(\phi^p_1-\phi^s_1)
+0.001 \sin(\phi^p_1-\phi^s_3) -0.0024 \sin(\phi^p_3-\phi^s_1)
\label{alambda}
\end{equation}

In the case of the minimal standard model, the effective weak
Hamiltonian in the notation of Buras \cite{buras} is,
\begin{equation}
H_W^{SM} = {G_F \over \sqrt{2}}V^*_{ud}V_{us}\sum_i
\biggl(z_i(\mu) - {V^*_{td}V_{ts} \over V^*_{ud}V_{us}} y_i(\mu)\biggr)
Q_i(\mu) +
{\rm ~h.~c.}
\label{effweak}
\end{equation}

The calculation of the weak phases would proceed
by evaluating the hadronic matrix
elements of the four-quark operators in Eq.~\ref{effweak} to obtain real
and imaginary parts for the amplitudes, schematically:
\begin{equation}
\matel{p \pi}{H_w^{eff}}{\Lambda^0}|^I_\ell = {\rm Re}M^I_\ell + i {\rm Im}
M^I_\ell .
\label{schematic}
\end{equation}
At present, however, we do not know how to compute the matrix elements so
we cannot actually implement this calculation.

For a simple estimate, we can take the real part of the matrix
elements from experiment (assuming that the measured amplitudes are real,
that is, that $\cp$ violation is small), and compute the imaginary
parts in vacuum saturation. This approach provides a conservative
estimate for the weak phases because the model calculation of the
real part of the amplitudes is smaller than the experimental value.
Nevertheless, the numbers should be viewed with great caution.

In the standard model, use of vacuum saturation to estimate the
matrix elements gives \cite{steger}
$\phi^1_s \approx -3 y_6 {\rm Im}\tau$ and
$\phi^1_p \approx -0.3 y_6 {\rm Im}\tau$.
Using $y_6 \approx -0.08$ \cite{buras}; and
${\rm Im}\tau = A^2 \lambda^4 \eta \leq 0.001$
we find \cite{latest}:
\begin{equation}
A(\Lambda^0_-) \approx 3 \times 10^{-5}
\label{vsnumres}
\end{equation}

Other models of $\cp$ violation contain additional short distance operators
with $\cp$ violating phases.
In Table~\ref{t: operators} we list all the four quark operators of
dimension six that change strangeness by one unit \cite{buchmuller}.

\begin{table}[tbh]
\centering
\caption[]{Dimension six $|\Delta S|=1$ four-quark operators.}
\begin{tabular}{|c|c|c|} \hline
Operator & Ref.~\cite{buchmuller} & $|\Delta S| =1$ \\ \hline
${\cal O}_{qq}^{(1,1)}$ & ${\cal O}^{(1,1)}_{}$ &
${1\over 2}\dd_L \gm s_L (\uu_L\gm u_L + \dd_L \gm d_L)$ \\
${\cal O}_{qq}^{(8,1)}$ & ${\cal O}^{(8,1)}_{}$ &
${1\over 2}\dd_L \ll \gm s_L (\uu_L\ll \gm u_L + \dd_L \ll\gm d_L)$ \\
${\cal O}_{qq}^{(1,3)}$ & ${\cal O}^{(1,3)}_{}$ &
${1\over 2}(2\uu_L \gm s_L \dd_L\gm u_L - \uu_L\gm u_L\dd_L\gm s_L +
\dd_L\gm d_L\dd_L\gm s_L  )$ \\
${\cal O}_{qq}^{(8,3)}$ & ${\cal O}^{(8,3)}_{}$ &
${1\over 2}(2\uu_L \ll\gm s_L \dd_L \ll\gm u_L - \uu_L \ll\gm
u_L\dd_L\ll\gm s_L+\dd_L\ll\gm d_L\dd_L\ll\gm s_L )$ \\ \hline
${\cal O}_{dd}^{(1)}$ & ${\cal O}^{(1)}_{dd}$ &
${1\over 2}\dd_R\gm s_R \dd_R \gm d_R$ \\
${\cal O}_{ud}^{(1)}$ & ${\cal O}^{(1)}_{ud}$ &
${1\over 2}\uu_R\gm u_R \dd_R \gm s_R$ \\
${\cal O}_{dd}^{(8)}$ & ${\cal O}^{(8)}_{dd}$ &
${1\over 2}\dd_R \ll\gm s_R \dd_R \ll \gm d_R$ \\
${\cal O}_{ud}^{(8)}$ & ${\cal O}^{(8)}_{ud}$ &
${1\over 2}\uu_R \ll\gm u_R \dd_R \ll\gm s_R$ \\ \hline
${\cal O}_{qu}^{(1)}$ & ${\cal O}^{(1)}_{qu}$ &
$\dd_L u_R \uu_R s_L$ \\
${\cal O}_{qu}^{(8)}$ & ${\cal O}^{(8)}_{qu}$ &
$\dd_L\ll u_R \uu_R \ll s_L$ \\
${\cal O}_{qd}^{(1)}$ & ${\cal O}^{(1)}_{qd}$ &
$\uu_L s_R\dd_R u_L +\dd_L s_R\dd_R d_L$ \\
${\cal O}_{qsd}^{(1)}$ & ${\cal O}^{(1)}_{qd}$ &
$\dd_L d_R \dd_R s_L$ \\
${\cal O}_{qsd}^{(8)}$ & ${\cal O}^{(8)}_{qd}$ &
$\dd_L\ll d_R \dd_R \ll s_L$ \\ \hline
${\cal O}_{qsq}^{(1)}$ & ${\cal O}^{(1)}_{qq}$ &
$-\uu_Rs_L\dd_Ru_L $ \\
${\cal O}_{qsq}^{(8)}$ & ${\cal O}^{(8)}_{qq}$ &
$-\uu_R \ll s_L\dd_R \ll u_L $ \\
${\cal O}_{qqs}^{(1)}$ & ${\cal O}^{(1)}_{qq}$&
$\dd_R s_L \uu_R u_L$ \\
${\cal O}_{qqs}^{(8)}$ & ${\cal O}^{(8)}_{qq}$&
$\dd_R\ll s_L \uu_R \ll u_L$ \\
${\cal O}_{qq}^{(1s)}$ & ${\cal O}^{(1)}_{qq}$&
$\uu_L u_R \dd_L s_R - \dd_L u_R \uu_L s_R $ \\
${\cal O}_{qq}^{(8s)}$ & ${\cal O}^{(8)}_{qq}$&
$\uu_L \ll u_R \dd_L\ll s_R - \dd_L\ll u_R \uu_L\ll s_R $
\\ \hline
\end{tabular}
\label{t: operators}
\end{table}

We estimate the contribution of these operators to $A(\Lambda^0_-)$
in vacuum saturation taking the real part of the amplitudes from
experiment as before \cite{latest}. The effective Hamiltonian now reads:
\begin{equation}
H_{eff} = H_W^{SM} + {g^2 \over \Lambda^2}
\biggl(\sum_i\lambda_i {\cal O}^{new}_i + {\rm ~h.~c.}\biggr)
\label{effh}
\end{equation}
The new operators violate $\cp$ if the coupling $\lambda_i$ has an imaginary
part. In this case they also contribute to $\cp$ violation in kaon decays.
For direct $\cp$ violation we write:
\begin{equation}
{\epsilon^\prime \over \epsilon}=\biggl({\epsilon^\prime \over \epsilon}
\biggr)_6 \biggl( 1 -\Omega_{SM}-\Omega_{NEW}\biggr).
\end{equation}
We require that $\Omega_{NEW}\leq 1$ to place bounds on the parity violating
$\cp$ violating phases using vacuum saturation to estimate the matrix elements.

In general, $\epsilon^\prime$ provides tighter constraints on new
$\cp$ violating interactions that does $\epsilon$. Nevertheless,
it is necessary to consider constraints from $\epsilon$ because
the ones that arise from $\epsilon^\prime$ do not apply to parity
conserving operators that do not contribute to the decay $K^0 \ra
\pi \pi$. Each of the new operators contributes to $\epsilon$:
\begin{equation}
|\epsilon|_i \approx {1 \over \sqrt{2}}{|{\rm Im}M_{12}|_i \over \Delta m_k}.
\end{equation}
We require that $|\epsilon|_i \leq |\epsilon|_{exp}$
to place bounds on the parity conserving $\cp$ violating
phases using the model of Ref.~\cite{dhlin} and of Ref.~\cite{dohowe}.

The bounds on the weak phases are presented in
Table~\ref{t: result} \cite{latest}.
The blank entries indicate that there is no bound because the particular
operator does not contribute to that amplitude.
\begin{table}[tbh]
\centering
\caption[]{Bounds on the phases that enter $A(\Lambda^0_-)$.}
\begin{tabular}{|c|c|c|c|c|} \hline
Operator & $\phi^p_1 \times 10^{5}$ & $\phi^s_1\times 10^{5}$
 & $\phi^p_3\times 10^{5}$ & $\phi^s_3\times 10^{5}$\\ \hline
${\cal O}_{qq}^{(1,1)}$  & $2.9$ & $-10$ & $--$ & $--$ \\
${\cal O}_{qq}^{(8,1)}$  & $9.0$ & $-10$ & $--$ & $--$ \\
${\cal O}_{qq}^{(1,3)}$  & $10$ & $-10$ & $--$ & $--$ \\
${\cal O}_{qq}^{(8,3)}$  & $-24$ & $-10$ & $--$ & $--$ \\ \hline
${\cal O}_{dd}^{(1)}$  & $5.3$ & $-10$ & $400$ & $-16$ \\
${\cal O}_{ud}^{(1)}$  & $-3.7$ & $-10$ & $400$ & $-16$ \\
${\cal O}_{dd}^{(8)}$  & $5.3$ & $-10$ & $400$ & $-16$ \\
${\cal O}_{ud}^{(8)}$  & $14$ & $-10$ & $400$ & $-16$ \\ \hline
${\cal O}_{qu}^{(1)}$  & $14$ & $-9.3$ & $400$ & $-15$ \\
${\cal O}_{qu}^{(8)}$  & $-24$ & $-10$ & $400$ & $-16$ \\
${\cal O}_{qd}^{(1)}$  & $9$ & $22$ & $--$ & $--$ \\
${\cal O}_{qsd}^{(1)}$  & $5.4$ & $2.8$ & $-400$ & $-15$ \\
${\cal O}_{qsd}^{(8)}$  & $5.3$ & $-10$ & $400$ & $-16$ \\ \hline
${\cal O}_{qsq}^{(1)}$  & $16$ & $-14$ & $400$ & $-15$ \\
${\cal O}_{qsq}^{(8)}$  & $24$ & $-2.8$ & $-400$ & $15$ \\
${\cal O}_{qqs}^{(1)}$  & $-74$ & $4.2$ & $400$ & $-15$ \\
${\cal O}_{qqs}^{(8)}$  & $14$ & $-9.3$ & $400$ & $-15$ \\
${\cal O}_{qq}^{(1s)}$  & $29$ & $22$ & $--$ & $--$ \\
${\cal O}_{qq}^{(8s)}$  & $29$ & $22$ & $--$ & $--$ \\ \hline
\end{tabular}
\label{t: result}
\end{table}

The bounds on the $p$-wave phases arise from the contributions of the
operator to $\epsilon$, and are weaker than the bounds on the $s$-wave
phases that arise from the contributions to $\epsilon^\prime$.
We illustrate separately the bounds on each parity
and isospin amplitude because it
is possible to construct operators with definite parity and isospin.

There are also two-quark operators that
can contribute to the processes under consideration \cite{buchmuller}:
\begin{equation}
{\cal O}_{dG} = (\overline{q}\sigma_{\mu\nu}\lambda^a d)\phi G^a_{\mu\nu}
\end{equation}
Following the procedure used for the four-quark operators but taking the
matrix elements from MIT bag model calculations this time \cite{dohowe}, we
find \cite{latest}:
\begin{equation}
A(\Lambda^0_-)  \leq  \cases{ 3 \times 10^{-4} & Parity conserving operator \cr
6 \times 10^{-5} & Parity violating operator \cr}
\label{respen}
\end{equation}

To summarize, the minimal standard model predicts
that $A(\Lambda^0_-)$ is of the order of a few times $10^{-5}$ \cite{steger}.
The effect of $\cp$
violating operators beyond the standard model is constrained by
the $\cp$ violation observed in kaon decays. Most of all the possible
dimension six operators
would naturally induce contributions to $A(\Lambda^0_-)$ at the $10^{-5}$
level, making them indistinguishable from the minimal standard model
(as long as precise calculations of the matrix elements are not available),
and inaccessible to the search to be conducted by E871. However, there
are certain operators, ${\cal O}_{qqs}^{(1)}$ and ${\cal O}_{dG}$ that
could induce an asymmetry $A(\Lambda^0_-)$ as large as a few times
$10^{-4}$.

The work described in this talk was done in collaboration with
Xiao~Gang~He and was supported in part by a DOE OJI award under contract
number DE-FG02-92ER40730.

\end{document}